\documentclass[showpacs,preprintnumbers]{revtex4}
\usepackage{graphicx}

\begin{document}

\title{Graphene-based one-dimensional photonic crystal}
\author{Oleg L. Berman and Roman Ya. Kezerashvili}
\affiliation{Physics
Department, New York City College
of Technology, The City University of New York,
Brooklyn, NY 11201, USA \\
and \\
The Graduate School and University Center, The
City University of New York, \\
New York, NY 10016, USA}

\begin{abstract}
A novel type of one-dimensional (1D) photonic crystal formed by the
array of periodically located stacks of alternating graphene and
dielectric stripes embedded into a background dielectric medium is
proposed. The wave equation for the electromagnetic wave propagating
in such a structure is solved in the framework of the Kronig-Penney
model. The frequency band structure of 1D graphene-based photonic
crystal is obtained analytically as a function of the filling factor
and the thickness of the dielectric between graphene stripes. The
photonic frequency corresponding to the electromagnetic wave
localized by the defect of photonic crystal formed by the extra
dielectric placed on the place of the stack of alternating graphene
and dielectric stripes is obtained.
\end{abstract}

\pacs{42.70.Qs, 78.67.Wj, 78.67.-n, 78.67.Pt}
\maketitle



\section{Introduction}

\label{intro}

Photonic crystals are formed by structures with the dielectric
constant periodically varying in space~\cite{Yablonovitch1}.
Electromagnetic waves in photonic crystals have a band spectrum and
a coordinate dependence caused by this periodicity of the dielectric
constant. Interestingly enough that the solution of Maxwell's
equations with a periodic dielectric constant,
resulting in photonic band-gap structures, is similar to the solution to Schr%
\"{o}dinger's equation for a periodic potential, resulting in the electron
energy band-gap structures in solids. Electromagnetic waves penetrate in the
photonic crystal similar to the Bloch waves of electrons in a regular
crystal. The width of the photonic band gap depends on the geometrical
parameters of the photonic crystal and the contrast of the dielectric
constants of the constituent elements~\cite{Joannopoulos1,Joannopoulos2}.
Different materials have been used for the corresponding constituent
elements including dielectrics, semiconductors and metals \cite%
{Joannopoulos1,Joannopoulos2,Sun_Jung,Sun,Maradudin,Kuzmiak}. The photonic
crystals with superconducting elements have been studied in Refs.~[%
\onlinecite{Takeda1,Takeda2,Zakhidov,Berman1,Lozovik,Berman2,Berman3}].The
one-dimensional photonic crystals formed by semiconductors were analyzed in
Ref.~[\onlinecite{Schulkin}]. It is well known that the properties of
photonic crystals provide an opportunity to manipulate the emission,
propagation and distribution of light~\cite{Lin1,Lin2} and photonic crystals
can be used as frequency filters. The properties of photonic crystals
were reviewed in Ref.~[\onlinecite{Wegener}].

Photonic crystals are different from the regular solid crystals in the
following way. While Schr\"{o}dinger's equation describes the regular solid
crystals via the scalar wave function, Maxwell's equations for photonic
band-gap crystals describes the electric or magnetic field, which is a
vector corresponding to the transverse electromagnetic waves.

A novel type of 2D electron system was experimentally obtained in graphene,
which is a 2D honeycomb lattice of the carbon atoms that form the basic
planar structure in graphite~\cite{Novoselov1,Zhang1}. Due to unusual
properties of the band structure, electronic properties of graphene became
the object of many recent experimental and theoretical studies \cite%
{Novoselov1,Zhang1, Novoselov2,Zhang2,Falko,Katsnelson,Castro_Neto_rmp}.
Graphene is a gapless semiconductor with massless electrons and holes which
have been described as Dirac-fermions \cite{DasSarma}. The unique electronic
properties of graphene in a magnetic field have been studied recently \cite%
{Nomura,Jain,Gusynin1,Gusynin2}. The space-time dispersion of graphene
conductivity was analyzed in Ref.~[\onlinecite{Varlamov}]. A graphene-based
two-dimensional photonic crystal was proposed in Ref. ~[%
\onlinecite{Berman_PLA}], where its frequency band structure was studied.
Different types of 2D photonic crystals were reviewed in Ref.~[%
\onlinecite{BBKL_review}].

In this Paper, we consider a one-dimensional photonic crystal formed by an
array of periodically located parallel stacks of alternating graphene
and dielectric stripes embedding into a background dielectric medium. The
graphene stripes are placed one under the other with the dielectric stripes
placed between them. We calculate the frequency band structure of such a
photonic crystal. The photonic band structure usually can be obtained using numerical calculations. In this Paper we obtain the
 analytical solution for the wave equation with the
periodical dielectric function. We also calculate the frequency
corresponding to the electromagnetic wave localized due to a defect
in the array of the stacks of graphene stripes separated by a
dielectric stripes.

The Paper is organized in the following way. In Sec.~\ref{sd} we obtain the
photonic band structure of the 1D graphene-based photonic crystal. In Sec.~%
\ref{defect} we find the frequency corresponding to the electromagnetic wave
localized by the defect of a 1D graphene-based photonic crystal. Finally, the
discussion of the results and conclusions follow in Sec.~\ref{disc}.


\section{The wave equation for 1D photonic crystal with graphene stripes}

\label{sd}

We consider  polarized electromagnetic waves with the electric field $%
\mathbf{E}$ perpendicular to the plane of graphene stripes. The wave
equation for the electric field in a dielectric media has the form~\cite%
{Landau}
\begin{eqnarray}  \label{we}
\Delta \mathbf{E}(\mathbf{r},t)-\frac{\varepsilon (\mathbf{r},t)}{c^{2}}\frac{%
\partial ^{2}\mathbf{E}(\mathbf{r},t)}{\partial t^{2}}=0\ ,
\end{eqnarray}
where $\varepsilon (\mathbf{r},t)$ is the dielectric constant of the media,
and $c$ is the speed of light in vacuum. Looking for solutions with harmonic
time variation of the electric field, i.e., $\mathbf{E}(\mathbf{r},t)=%
\mathbf{E}(\mathbf{r})e^{i\omega t}$, and considering the propagation of
wave in the $x-$direction along the plane of graphene stripes and
perpendicular to the graphene-dielectric boundaries one obtains from Eq.~(%
\ref{we})
\begin{eqnarray}  \label{wex}
\frac{\partial ^{2}E_{z}(x)}{\partial x^{2}}+\frac{\omega
^{2}\varepsilon (x,\omega )}{c^{2}}E_{z}(x)=0\ .
\end{eqnarray}

\begin{figure}[tbp]
\includegraphics[width = 3.5in]{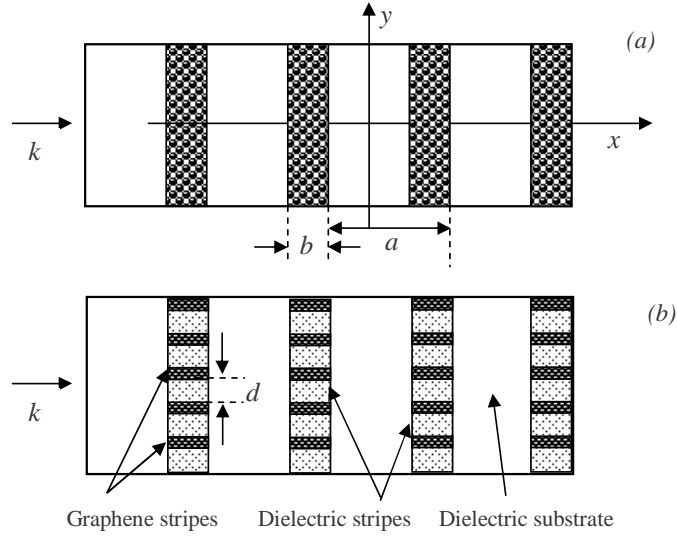}
\caption{The 1D photonic crystal with graphene stripes. (a) the top
view; (b) the side view. The material of the dielectric stripes
between graphene stripes can be the same as the material of the
dielectric substrate.} \label{pc}
\end{figure}

The dielectric constant of the 1D periodic structure is given by
\begin{eqnarray}
\varepsilon (x,\omega )=\left\{%
\begin{array}{c}
\varepsilon _{0},\text{ }\mathrm{for}\ \ \ -\frac{1}{2}(a-b)+na<x<\frac{1}{2}%
(a-b)+na\ , \\
\varepsilon _{1}(\omega ),\text{ \ }\mathrm{for}\ \ \ \frac{1}{2}(a-b)+na<x<%
\frac{1}{2}(a+b)+na,%
\end{array}%
\right.  \nonumber  \label{model}
\end{eqnarray}
where $\varepsilon _{0}$ is the dielectric constant of the dielectric, $%
\varepsilon _{1}(\omega)$ is the dielectric function of graphene
multilayers separated by the dielectric barriers, $a$ is the period
of 1D array of graphene stripes, $b$ is the width of graphene
stripes, and $n$ is an integer. The 1D photonic crystal with
graphene stripes separated by dielectric layers with a thickness $d$
is shown in Fig.~\ref{pc}. By introducing the filling factor $f$ the
relation between $a$ and $b$ can be written as $b=af$.

The dielectric function $\varepsilon _{1}(\omega )$ of a graphene multilayers
system separated by dielectric layers with a dielectric constant $%
\varepsilon _{0}$ and a thickness $d$ is given by~\cite%
{Falkovsky_prb,Falkovsky_conf}
\begin{eqnarray}  \label{dielmult}
\varepsilon _{1}(\omega )=\varepsilon _{0}+\frac{4\pi i\sigma _{g}(\omega )}{%
\omega d}\ ,
\end{eqnarray}
where $\sigma _{g}(\omega )$ is the dynamical conductivity of doped graphene
for the high frequencies ($\omega \gg kv_{F}$, $\omega \gg \tau ^{-1}$) at
temperature $T$ given by~\cite{Falkovsky_prb,Falkovsky_conf}
\begin{eqnarray}  \label{cond}
&&\sigma _{g}(\omega )=\frac{e^{2}}{4\hbar }\left[ \eta (\hbar \omega -2\mu
)\right.  \nonumber \\
&&\left. +\frac{i}{2\pi }\left( \frac{16k_{B}T}{\hbar \omega }\log \left[
2\cosh \left( \frac{\mu }{2k_{B}T}\right) \right] \right. \right.  \nonumber
\\
&&\left. \left. -\log \frac{(\hbar \omega +2\mu )^{2}}{(\hbar \omega -2\mu
)^{2}+(2k_{B}T)^{2}}\right) \right] \ .
\end{eqnarray}

Here $e$ is the charge of an electron, $\tau ^{-1}$ is the electron
collision rate, $\eta $ is the Heaviside step function, $k_{B}$ is the
Boltzmann constant, and $\mu $ is the the chemical potential determined by
the electron concentration, which is controlled by the doping. The chemical
potential can be calculated as $\mu =(\pi n_{0})^{1/2}\hbar v_{F}$, where the
electron concentration is given by $n_{0}=(\mu /(\hbar v_{F}))^{2}/\pi $ and
$v_{F}=10^{8}\ \mathrm{cm/s}$ is the Fermi velocity of electrons in graphene~%
\cite{Falkovsky_conf}.

We solve the wave equation (\ref{wex}) to find the eigenfrequencies
corresponding to the electromagnetic wave penetrating in the photonic
crystal shown in Fig.~\ref{pc}. This wave equation is mathematically similar
to the Schr\"{o}dinger equation for an electron moving in a one-dimensional
rectangular periodic potential barrier described by the Kronig-Penney
model. The eigenenergies of the Schr\"{o}dinger equation corresponding to the
Kronig-Penney model presented in Eq.~(\ref{modelsch}) are given by~\cite%
{Kronig}:
\begin{eqnarray}  \label{en}
\cos (ka)=\cosh (\beta b)\cos \left[ \alpha a(1-f)\right] +\frac{\alpha
^{2}-\beta ^{2}}{2\alpha \beta }\sinh (\beta b)\sin \left[ \alpha a(1-f)%
\right] \ ,
\end{eqnarray}
where the wave vector $k$ is in the range $0\leq k\leq 2\pi /a$, and $\alpha
$ and $\beta $ are defined as
\begin{eqnarray}  \label{ab}
\alpha &=&\frac{\sqrt{\varepsilon _{0}}}{c}\omega \ ,  \nonumber \\
\beta &=&\frac{\sqrt{\varepsilon _{1}(\omega )}}{c}\omega \ .
\end{eqnarray}%
%
%
Applying Eq.~(\ref{ab}), Eq.~(\ref{en}) can be written in a form
\begin{eqnarray}  \label{en2}
\cos (ka)=\cosh \left( \sqrt{\varepsilon _{1}(\omega )}\frac{af}{c}\omega
\right) \cos \left[ \sqrt{\varepsilon _{0}}a\frac{(1-f)}{c}\omega \right] +%
\frac{\varepsilon _{0}-\varepsilon _{1}(\omega )}{2\sqrt{\varepsilon
_{0}\varepsilon _{1}}}\sinh \left( \sqrt{\varepsilon _{1}(\omega )}\frac{af}{%
c}\omega \right) \sin \left[ \sqrt{\varepsilon _{0}}\frac{a(1-f)}{c}\omega %
\right] \ .
\end{eqnarray}%
%
%
The eigenfrequencies of the 1D photonic crystal as functions of the
wave vector $k$ can be obtained by substituting the dielectric
constant of the multilayer graphene given by Eq. (3)~into
Eq.~(\ref{en2}). The solutions of Eq.~(\ref{en2}) for $\omega $ as
functions of $k$ provides the frequency band structure for 1D
photonic crystal formed by the periodically located parallel stacks
of alternating graphene and dielectric stripes embedded into a
background dielectric medium and allow one to determine  the
photonic band gap $\Omega $. Eq.~(\ref{en2}) shows that the photonic
band gap depends on the filling factor $f$ and the period of 1D
array of graphene stripes $a$, as well as on the thickness $d$ of
the dielectric stripes that separate the graphene stripes. Let us
mention that dependence of the photonic band gap on the thickness
$d$ of the dielectric layer, which separates the graphene stripes,
is presented in (\ref{en2}), since the dielectric function
$\varepsilon_{1}(\omega)$~(\ref{dielmult}) of a graphene multilayers
system separated by dielectric layers depends on $d$.

\section{Localization of the electromagnetic wave on the defect}

\label{defect}

Let us consider a defect in the array of stacks of alternating
graphene and dielectric stripes embedded into a background
dielectric medium. This defect is formed by one empty space or
\textquotedblleft 1D vacancy\textquotedblright\ due to the absence
of a stack of alternating graphene and dielectric stripes in one
place, where it should be placed due to the periodicity. This place
is filled by the dielectric. This extra dielectric stipe contributes
to the dielectric contrast that results by adding the term $-\omega
^{2}/c^{2}(\varepsilon _{1}(\omega )-\varepsilon _{0})\gamma
(|x-x_{0}|)E_{z}(x)$ to the r.h.s. in Eq.~(\ref{wex}). Here $\gamma
(|x-x_{0}|)=1$ for $|x-x_{0}|\leq b$, and $\gamma (|x-x_{0}|)=0$ for
$|x-x_{0}|>b$, where $x_{0}$ corresponds to the coordinate in the
middle of 1D defect, which is the coordinate of the middle of the
absent graphene stripe. As a result we obtain the wave equation for
the electric field for 1D photonic crystal with the defect:
\begin{eqnarray}
\label{realcrys}
\frac{\partial ^{2}E_{z}(x)}{\partial x^{2}}+\frac{\omega ^{2}}{%
c^{2}}\left( \varepsilon (x,\omega )-(\varepsilon _{1}(\omega
)-\varepsilon _{0})\gamma (|x-x_{0}|)\right) E_{z}(x)=0\ .
\end{eqnarray}
Eq.~(\ref{realcrys}) describes the periodic array of stacks of alternating
graphene and dielectric stripes with the defect formed by one stack of
alternating graphene and dielectric stripes being absent.

We solve the wave equation (\ref{realcrys}) to find the eigenfrequency
corresponding to the electromagnetic wave localized at the defect formed by
a background dielectric medium due to the absence of the stack of
alternating graphene and dielectric stripes in one place, where it should be
placed due to the periodicity. This wave equation is similar to the Schr\"{o}%
dinger equation describing the electron in the periodic potential energy in
the presence of the potential energy of the defect placed at the point $%
x_{0} $ given in Appendix B. From the mapping of Klein-Gordon type equation
given by Eq.~(\ref{kgsim}) the wave equation for the electric field in
graphene-based photonic crystal has the following form
\begin{eqnarray}  \label{eigen_eq}
-\frac{4k_{0}^{2}c^{4}}{3}\frac{d^{2}E_{z}(x)}{dx^{2}}-2 \omega
^{2}\Omega^{2}(\varepsilon _{1}(\omega )-\varepsilon _{0})\gamma
(|x-x_{0}|)E_{z}(x)=(\omega ^{4}-\Omega ^{4})E_{z}(x)\ ,
\end{eqnarray}
where $k_{0} = 2\pi/a$ is the vector of the 1D reciprocal lattice.
 This Klein-Gordon type equation has the eigenvalue $\omega
^{4}-\Omega ^{4}$. In Eq.~(\ref{eigen_eq}) $\Omega $ is the width of the forbidden band
(photonic gap) in the spectrum of the electromagnetic wave.

The electric field can be obtained by mapping Eqs.~(\ref{eigen_f111})
and~(\ref{eigen_f00}) that correspond to wavefunctions of the
stationary states. The continuity of the
electric field and its derivative in the points $x=x_{0}+b/2$ and $%
x=x_{0}-b/2$ result in the transcendental equation determining the spectrum
of the even states:
\begin{eqnarray}  \label{eigen_v}
\sqrt{\omega ^{2}\Omega^{2}(\varepsilon _{1}(\omega )-\varepsilon
_{0})^{2}-\omega ^{4}+\Omega ^{4}}\tan \left( \sqrt{\omega
^{2}\Omega^{2}(\varepsilon _{1}(\omega
)-\varepsilon _{0})^{2}-\omega ^{4}+\Omega ^{4}}b^{2}/c^{2}\right) =\sqrt{%
|\omega ^{4}-\Omega ^{4}|}\ ,
\end{eqnarray}
as well as of the odd states:
\begin{eqnarray}  \label{eigen_v0}
\sqrt{\omega ^{2}\Omega^{2}(\varepsilon _{1}(\omega )-\varepsilon
_{0})^{2}-\omega ^{4}+\Omega ^{4}}\cot \left( \sqrt{\omega
^{2}\Omega^{2}(\varepsilon _{1}(\omega
)-\varepsilon _{0})^{2}-\omega ^{4}+\Omega ^{4}}b^{2}/c^{2}\right) =-\sqrt{%
|\omega ^{4}-\Omega ^{4}|}\ .
\end{eqnarray}

Solving Eqs.~(\ref{eigen_v}) and~(\ref{eigen_v0}) with respect to the
frequency $\omega$, we obtain the frequency corresponding to the
photonic mode localized by the defect.

\section{Discussion}

\label{disc}

\begin{figure}[ht]
\includegraphics[width=0.9\textwidth]{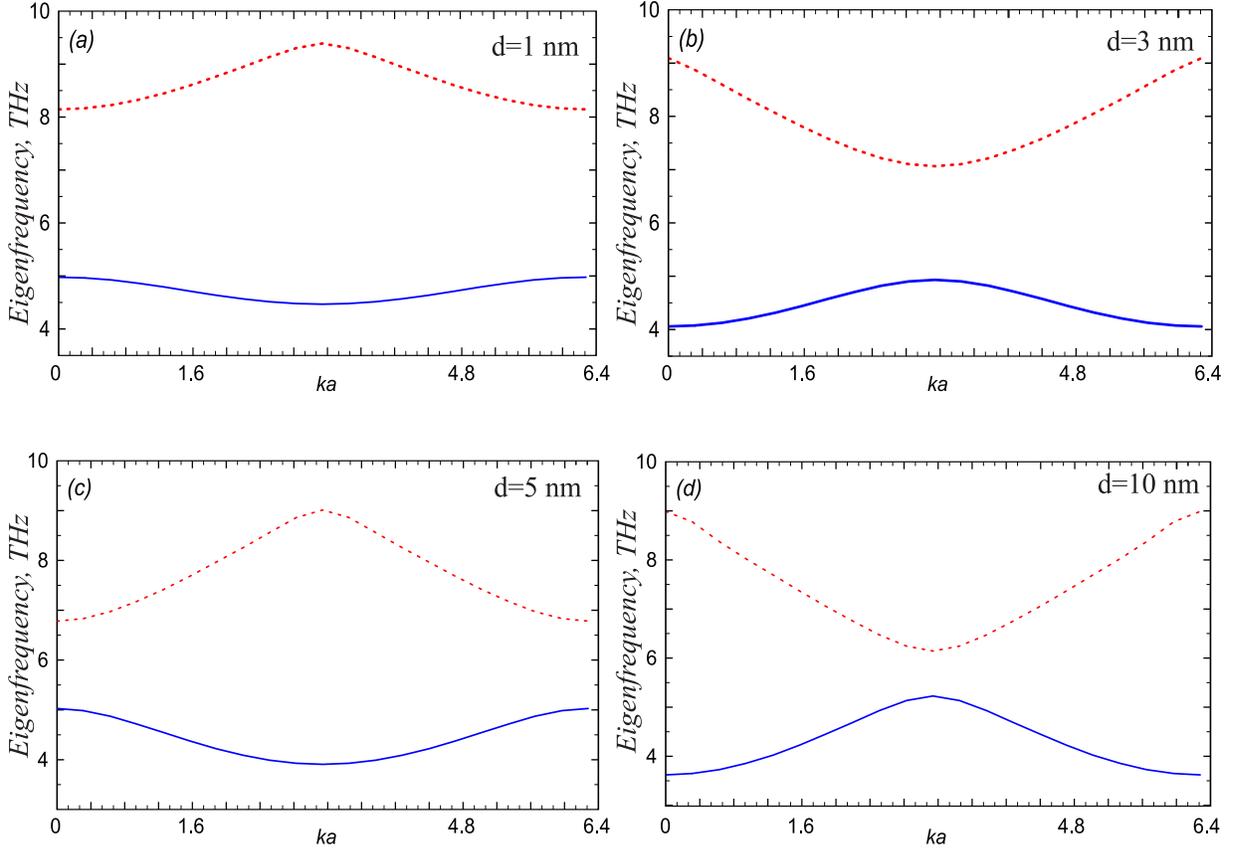}
\caption{The dispersion relation for the 1D graphene-based photonic crystal with  $a=25 \ \mathrm{\mu m}$ and
filling factor $f=0.3927$ for  different  thicknesses  of the
dielectric between graphene stripes.}
\label{pbs}
\end{figure}

In the calculations below we assume $n_{0}=10^{11}\ \mathrm{cm^{-2}}$. For
simplicity, we consider the same material for a background dielectric medium
and dielectric stripes between the graphene stripes. As the dielectric
material we consider SiO$_{2}$ with the dielectric constant $\varepsilon
_{0}=4.5$. Using Eq.~(\ref{en2}) we calculated the band structure for 1D graphene-based photonic crystal.
 The results of the calculations of the dispersion relation of the photonic crystal are presented in Fig. \ref{pbs}. The
photonic band structure is calculated for different distances between
graphene layers $d$. In our calculations we used the chemical potential for
the electrons in graphene $\mu =3.525\times 10^{-21}J$, temperature $T=300\
\mathrm{K}$, the period of the 1D graphene stripes array $a=25\times
10^{-6}\ \mathrm{m}$, and the filling factor $f=0.3927$. According to the
results of our calculations, the photonic band structure almost does not
depend on $\varepsilon _{0}$ due to the fact that $\varepsilon _{0}\ll
|\varepsilon _{1}|$. The results of our calculations demonstrate the strong
dependence of the photonic band structure on the thickness  $d$ of the dielectric that separates the graphene stripes. At $d=1\ \mathrm{nm}$ and $%
d=5\ \mathrm{nm}$ the distance between the lower and the upper dispersion
curves is larger in the middle than at the edges. At $d=3\ \mathrm{nm}$ and $%
d=10\ \mathrm{nm}$ the distance between the lower and the upper dispersion
curves is larger at the edges than in the middle.

Using Eqs.~(\ref{eigen_v}) and~(\ref{eigen_v0}) we calculate
frequencies corresponding to the electromagnetic wave localized by
the defect in the photonic crystal formed due to the absence of a
stack of alternating graphene and dielectric stripes in one place,
where it should be placed due to the periodicity. For the photonic
crystal  with the defect at $d=1\ \mathrm{nm}$ we have $\nu =1.79\
\mathrm{THz}$, at $d=3\ \mathrm{nm}$ we have $\nu =1.85\
\mathrm{THz}$, at $d=5\ \mathrm{nm}$ we have $\nu =3.91\
\mathrm{THz}$, at $d=10\ \mathrm{nm}$ we have $\nu =3.62\
\mathrm{THz}$. All these frequencies are located inside the photonic
band gap. Let us mention that since our approach is based on the
Luttinger-Kohn model \cite{Luttinger,Kohn}, the localized frequency
$\nu$ does not depend on momentum. The frequency localized by the
defect is just a constant in the photonic band structure, which can
be controlled by the thickness $d$ of the dielectric stripes that
separate the graphene stripes, as well as by the filling factor $f$
and the period of 1D array of graphene stripes $a$.

In conclusion, the graphene-based photonic crystal proposed in this
paper is a novel type one-dimensional photonic crystal formed by the
array of the stacks of alternating graphene and dielectric stripes
embedded into a background dielectric medium. This system can be
analyzed as a 1D photonic crystal and the corresponding wave
equation for the electromagnetic wave propagating in such a
structure can be solved in the framework of the Kronig-Penney model.
The frequency band structure is obtained analytically as a function
of the filling factor, the period of 1D array of graphene stripes
and the thickness of the dielectric between the graphene stripes. We
obtain the photonic frequency corresponding to the electromagnetic
wave localized at the defect of a photonic crystal formed by a
background dielectric medium due to the absence of the stack of
alternating graphene and dielectric stripes in one place, where it
should be placed due to the periodicity.





\acknowledgments

This research was supported by PSC CUNY grant: Award \# 64197-00 42.

\appendix


\section{The Schr\"{o}dinger equation with the periodic potential}

\label{ap.map}

The wave equation~(\ref{wex}) with the dielectric constant given by Eq.~(\ref%
{model}) can be mapped onto the Schr\"{o}dinger equation for an electron in
the 1D periodic potential function approximated by a rectangular potential
barrier, used in the Kronig-Penney model~\cite{Kronig}:
\begin{eqnarray}
-\frac{\hbar ^{2}}{2m_{0}}\frac{\partial ^{2}\psi (x)}{\partial x^{2}}
&=&E\psi (x)\ ,\hspace{3cm}\mathrm{for}\ \ \ -\frac{1}{2}(a-b)+na<x<\frac{1}{%
2}(a-b)+na,  \nonumber  \label{modelsch} \\
-\frac{\hbar ^{2}}{2m_{0}}\frac{\partial ^{2}\psi (x)}{\partial x^{2}}+V(x)
&=&E\psi (x)\ ,\text{ }V(x)=\frac{\hbar ^{2}\omega ^{2}}{2m_{0}c^{2}}%
\varepsilon (x,\omega )\text{ }\mathrm{for}\ \frac{1}{2}(a-b)+na<x<\frac{1}{2%
}(a+b)+na,
\end{eqnarray}%
%
%
%
%
where $m_{0}$ is the mass of an electron, $\psi (x)$ and $E$ are the wave
function and energy of an electron, $V(x)$ is the potential due to the ion
of the 1D crystal lattice approximated by a rectangular barrier, $a$ is the
period of 1D array of the scatterers with the rectangular potential, $b$ is
the width of the rectangular potential barrier, and $n$ is an integer.

It is easy to see that the wave equation~(\ref{wex}) with the dielectric
constant given by Eq.~(\ref{model}) can be mapped onto the Schr\"{o}dinger
equation~(\ref{modelsch}) for an electron in the 1D periodic potential
corresponding to the Kronig-Penney model. The mapping relations are the following%
%
%
%
%
\begin{eqnarray}
\psi (x) &\equiv &E_{z}(x)\ ,  \nonumber  \label{map} \\
\alpha ^{2} &=&\frac{2m_{0}}{\hbar ^{2}}E\equiv \frac{\varepsilon _{0}}{c^{2}%
}\omega ^{2}, \\
\beta ^{2} &=&\frac{2m_{0}}{\hbar ^{2}}(E-V(x))\equiv \frac{\varepsilon _{1}(\omega)%
}{c^{2}}\omega ^{2}.  \nonumber
\end{eqnarray}%
%
%
%
%
The solution of Eq. (\ref{modelsch}) is well known and is given by
expression (\ref{en}).


\section{The Dirac-type equations for the
the electromagnetic wave in the photonic crystal with the defect}

\label{solution}

\bigskip\ Following Ref.~[\onlinecite{Berman2}] we can map the wave equation
for the photonic crystal with the defect onto the Schr\"{o}dinger equation
in the periodic field with the defect. After mapping of Eq.~(\ref{realcrys})
onto the Schr\"{o}dinger equation describing the electron with the effective
electron mass $m_{0}$ in the periodic potential energy $V(x)$ in the
presence of the potential energy of the defect $W\gamma (|x-x_{0}|)$ placed
at the point $x_{0}$ we have
\begin{eqnarray}\label{sch1}
\left[ -\frac{\hbar ^{2}}{2m_{0}}\frac{d^{2}}{dx^{2}}+V(x)-W\gamma
(|x-x_{0}|)\right] \psi (x)=\varepsilon _{\omega }\psi (x)\ .
\end{eqnarray}
In Eq.(\ref{sch1}) $\psi (x)=E_{z}(x)$,
\[
\varepsilon _{\omega }=\frac{\hbar ^{2}\omega ^{2}}{2m_{0}c^{2}}\ ,
\]%
%
and the potential $V(x)$ is given in Eq. (A1) and potential $W$ is
defined as
\begin{eqnarray}
\label{imppot}
 W=\frac{\hbar ^{2}\omega
^{2}}{2m_{0}c^{2}}(\varepsilon _{1}(\omega )-\varepsilon _{0})\ .
\end{eqnarray}

Eq.~(\ref{sch1}) has the same form as Eq.~(6) in Ref.~[\onlinecite{Keldysh}%
]. However, in our case the potential $W$ is defined by Eq.~(\ref{imppot})
and corresponds to the potential of the defect in the Schr\"{o}dinger
equation for an \textquotedblleft electron\textquotedblright\ in the
periodic field of the crystal lattice and in the presence of the defect.

We will reduce the problem of the Schr\"{o}dinger equation for a
particle in the periodic potential $V(x)$ related to the system of
the periodically placed stacks of alternating graphene and
dielectric stripes embedded into a background dielectric medium and
an \textquotedblleft defect potential\textquotedblright\ $W$ related
to a defect to a much simpler equation for the envelope
wavefunctions. Taking into account the two-band structure and
following the two-band model~\cite{Keldysh} we introduce a
two-component spinor wave function $\psi (x)$
\begin{eqnarray}  \label{2comp}
\psi (x)=\left(
\begin{array}{c}
\varphi (x) \\
\chi (x)%
\end{array}%
\right) \ ,
\end{eqnarray}
where two different neighboring bands are described by wave
functions $\varphi (x)$ and $\chi (x)$. Note that Eq.~(\ref{sch1})
contains both the periodic function $V(x)$ corresponding to the ideal lattice and $W$, describing the potential of a defect. Without a defect the energy spectrum
would be described by two neighboring bands and the gap between them.
Applying the standard two-band approach, we  obtain an
effective Dirac-type equation for the envelope spinor wave function, which
implies the periodicity provided by $V(x)$~\cite{Keldysh,Berman2}.

Note that Eq.~(\ref{sch1}) describes an electron in the periodic potential
of the ideal crystal lattice $V(x)$ and the potential of the defect $W$. If
the solution corresponding to the absence of impurity $W=0$ is known, the
energy levels of the electron localized by the defect can be obtained by
replacing Eq.~(\ref{sch1}) by the Dirac-type equation according to
Luttinger-Kohn model described in Refs.~[\onlinecite{Luttinger,Kohn,Keldysh}%
]. This model implies a Dirac-type equation for the two-component spinor wave
function. According to Ref.~\cite{Keldysh}, the functions $\varphi _{n}(x)$
and $\chi _{n}(x)$ defined as
\begin{eqnarray}
\varphi _{n}(x) &=&\sum_{k}c_{n}(k)\exp \left[ ikx/\hbar \right] \ ,
\nonumber  \label{phid} \\
\chi _{n}(x) &=&\sum_{k^{\prime }}d_{n}(k^{\prime })\exp \left[ ik^{\prime
}x/\hbar \right]
\end{eqnarray}%
%
%
satisfy the set of the second order partial differential equations.
Considering only two neighboring bands corresponding to the wave function $%
\psi (x)$ given by Eq.~(\ref{2comp}) this set of equations for $\varphi
_{n}(x)$ and $\chi _{n}(x)$ can be reduced to the Dirac-type equations for
the two-component spinor~(\ref{2comp}), which has the following form \cite%
{Keldysh}
\begin{eqnarray}
\left[ \varepsilon _{\omega }-\Delta _{\omega }-W\gamma
(|x-x_{0}|)\right] \varphi (x)+i\hbar s \mathbf{\sigma
}_{x}\frac{d\chi (x)}{dx}
&=&0\ ,  \nonumber  \label{dir} \\
\left[ \varepsilon _{\omega }+\Delta _{\omega }-W\gamma
(|x-x_{0}|)\right] \chi (x)+i\hbar s \mathbf{\sigma
}_{x}\frac{d\varphi (x)}{dx} &=&0\ ,
\end{eqnarray}%
%
%
%
where $s = \hbar k_{0}/(\sqrt{3}m_{0})$. In Eq. (\ref{dir})
$\mathbf{\sigma }_{x}$ is Pauli matrix and $\Delta _{\omega }$ is
the width of the forbidden band in the electron spectrum that, as it
follows from the mapping of the wave equation for the electric field
graphene-based photonic crystal onto Eq.~(\ref{sch1}), is $\Delta _{\omega }=%
\frac{\hbar ^{2}\Omega ^{2}}{2m_{0}c^{2}},$ where $\Omega $ is the
width of the forbidden band (photonic gap) in the spectrum of the
electromagnetic wave. Eqs. (\ref{dir}) are obtained in the limit
$|\varepsilon _{\omega }^{2}-\Delta _{\omega }^{2}|/(2\Delta
_{\omega }^{2})\ll 1$. \
Defining the effective mass of a quasiparticle as $m_{\omega
}=3m_{0}^{2}\Delta _{\omega }/(\hbar ^{2}k_{0}^{2})\ ,$ and
following the standard procedure of quantum electrodynamics
\cite{Fermi,Bjorken} we obtain from the system of Dirac-type
Eqs.~(\ref{dir}) the following Klein-Gordon type equation. Note that the Klein Gordon equation can be reduced to a
Schr\"{o}dinger-like equation with an effective energy and an
effective potential. If the potential is weak enough to ignore the
$W^2$ term, the relativistic formalism becomes equivalent to the
non-relativistic formalism. More importantly, in situations where
the Klein-Gordon equation is not exactly solvable, the
Schr\"{o}dinger form of the Klein-Gordon equation sheds some light
on the problem as it could be reduced to a solvable eigenvalue
problem:
\begin{eqnarray}
\label{kgsim} \left[ -\hbar^{2}s^{2}\frac{d^{2}}{dx^{2}} -2m_{\omega
}s^{2}W\gamma (|x-x_{0}|)\right] \Psi (x)=\left(\varepsilon _{\omega
}^{2}-\Delta_{\omega }^{2}\right)\Psi (x)\ .
\end{eqnarray}

This Klein-Gordon type equation has the form of the 2D Schr\"{o}dinger
equation for a particle in the square well potential well with the
eigenvalue
\[
\mathcal{E}_{\omega }= \frac{\varepsilon _{\omega
}^{2}-\Delta_{\omega }^{2}}{2m_{\omega} s^{2}}  \ .
\]%
%
The wavefunctions of the even stationary states have the form ($\mathcal{E}%
_{\omega }<0$):
\begin{eqnarray}
\Psi (x) &=&A_{1}\cos \left[ \sqrt{2m_{\omega }(W-\mathcal{E}_{\omega
})/\hbar ^{2}}x\right] ,\ \ \ \ \ \ \ |x-x_{0}|\leq b\ ,  \nonumber
\label{eigen_f111} \\
\Psi (x) &=&B_{1}\exp \left[ -\sqrt{2m_{\omega }\mathcal{E}_{\omega }/\hbar
^{2}}x\right] ,\ \ \ \ \ \ \ |x-x_{0}|>b\ .
\end{eqnarray}%
%
%
The continuity of the wave function and its derivative in the points $%
x=x_{0}+b/2$ and $x=x_{0}-b/2$ result in the transcendental equation
determining the spectrum of the even quantum states:
\begin{eqnarray}\label{eigen_v111}
\sqrt{W-\mathcal{E}_{\omega }}\tan \sqrt{2m_{\omega }(W-\mathcal{E}_{\omega
})b^{2}/\hbar ^{2}}=\sqrt{|\mathcal{E}_{\omega }|}\ .
\end{eqnarray}

The wavefunctions of the odd stationary states have the form ($\mathcal{E}%
_{\omega }<0$):
\begin{eqnarray}
\Psi (x) &=&A_{2}\sin \left[ \sqrt{2m_{\omega
}(W-\mathcal{E}_{\omega })/\hbar ^{2}}x\right] ,\ \ \ \ \ \ \
|x-x_{0}|\leq b\ ,  \nonumber
\label{eigen_f00} \\
\Psi (x) &=&B_{2}\exp \left[ -\sqrt{2m_{\omega }\mathcal{E}_{\omega }/\hbar
^{2}}x\right] ,\ \ \ \ \ \ \ |x-x_{0}|>b\ .
\end{eqnarray}%
%
%
The continuity of the wave function and its derivative in the points $%
x=x_{0}+b/2$ and $x=x_{0}-b/2$ result in the transcendental equation
determining the spectrum of the odd quantum states:
\begin{eqnarray}\label{eigen_v00}
\sqrt{W-\mathcal{E}_{\omega }}\cot \sqrt{2m_{\omega }(W-\mathcal{E}_{\omega
})b^{2}/\hbar ^{2}}=-\sqrt{|\mathcal{E}_{\omega }|}\ .
\end{eqnarray}%

\end{document}